\documentclass[prb,twocolumn,showpacs,preprintnumbers]{revtex4}
\usepackage{graphicx}
\usepackage{amsfonts}
\usepackage{amssymb}
\usepackage{amsmath}
\usepackage{makeidx}
\makeindex
\newcommand{\be}{\begin{equation}}
\newcommand{\ee}{\end{equation}}
\newcommand{\bea}{\begin{eqnarray}}
\newcommand{\eea}{\end{eqnarray}}
\newcommand{\Eq}[1]{Eq.\,(\ref{#1})}
\newcommand{\Fig}[1]{Fig.\,\ref{#1}}
\newcommand{\Sec}[1]{Sec.\,\ref{#1}}
\newcommand{\Onlinecite}[1]{Ref.\,\onlinecite{#1}} 

\newcommand{\GFk}{\hat{\mathbf{G}}_k}
\newcommand{\GFPk}{\hat{\mbox{\boldmath ${\cal G}$}}\hspace*{-0.5pt}_k}
\newcommand{\Qn}{\hat{\mathbf{Q}}_n}
\newcommand{\br}{\mathbf{r}}

\newcommand{\En}{\mathbf{E}_n}
\newcommand{\Em}{\mathbf{E}_m}
\newcommand{\Fn}{\mathbf{F}_n}
\newcommand{\Epn}{\mbox{\boldmath ${\cal E}$}\hspace*{-1.5pt}_\nu}

\newcommand{\za}{\hat{\mathbf{z}}}

\newcommand{\brho}{\mbox{\boldmath ${\rho}$}}

\begin{document}
\title{Resonant state expansion applied to two-dimensional open optical systems}
\author{ M.\,B. Doost}
\affiliation{School of Physics and Astronomy, Cardiff University, Cardiff CF24 3AA,
United Kingdom}
\author{W. Langbein}
\affiliation{School of Physics and Astronomy, Cardiff University, Cardiff CF24 3AA,
United Kingdom}
\author{E.\,A. Muljarov}
\altaffiliation[]{egor.muljarov@astro.cf.ac.uk; on leave from General Physics Institute RAS,
Moscow, Russia.} \affiliation{School of Physics and Astronomy, Cardiff University, Cardiff CF24 3AA,
United Kingdom}
\begin{abstract}
The resonant state expansion (RSE), a rigorous perturbative method in electrodynamics, is applied
to two-dimensional open optical systems. The analytically solvable homogeneous dielectric cylinder
is used as unperturbed system, and its Green's function is shown to contain a cut in the complex
frequency plane, which is included in the RSE basis. The complex eigenfrequencies of modes are calculated using the RSE for a selection of perturbations which mix unperturbed modes of different orbital momentum, such as half-cylinder, thin-film and thin-wire perturbation, demonstrating the accuracy and convergency of the method. The resonant
states for the thin-wire perturbation are shown to reproduce an approximative analytical solution.
\end{abstract}
\date{\today}
\maketitle
\section{Introduction}
The electro-magnetic spectrum of an open optical system is characterized by its resonances.
Optical resonators,\cite{Vahala03} such as planar microcavities,\cite{Chang96} photonic crystal
fibers,\cite{Russel03} dielectric micro-spheres,\cite{Collot93}
micro-disks\cite{McCall92,Gayral99} and
micro-cylinders\cite{Chantada08,Dubertrand08,Dettmann09EPL} are examples of such systems designed
to have a series of narrow resonances in their optical spectra. Such resonances, known as cavity
modes in planar microcavities and whispering gallery modes (WGMs) in micro-spheres and -cylinders,
are characterized by their spectral positions and linewidths,  given, respectively, by the real and imaginary
part of the complex eigenfrequencies of the system.

Finite linewidths of resonances are typical for open systems and are due to energy leakage from
the system to the outside. This leakage can be enhanced by various structural imperfections and scatterers.
In particular, when an object is placed inside or in close proximity to
the cavity, the resulting modification of the electromagnetic susceptibility perturbs the cavity
resonances, changing both their position and linewidth, most noticeably for the high-quality (i.e.
narrow-line) resonances. This effect is the basis for resonant optical
biosensing.\cite{Vollmer08,Lutti08,Chantada08} The changes in the spectral properties of
resonators in the presence of perturbations can be used to characterize the size and shape of the
attached nanoparticles.\cite{Zhu10,Kippenberg10} The WGM resonances in microdisks and spherical
microcavities have been used in sensors for the characterization of nanolayers\cite{Noto05},
DNA\cite{Vollmer03} and protein molecules,\cite{Vollmer02} as well as for single
atom\cite{Rosenblit04} and nanoparticle detection.\cite{Shopova11,He11}
Other applications of high-quality modes 
include miniature laser sources\cite{Frateschi95,Sandoghdar96,Ilchenko06,Wang10} and
photonic-crystal optical fibers.\cite{Russel03}



While the eigenmodes of resonators with simple and highly symmetric geometries can in some cases
be calculated exactly, determining their perturbations presents a significant challenge as the
popular computational techniques in electrodynamics, such as the finite difference in time
domain\cite{Taflove00, Hagness97} or finite element method\cite{Wiersig03, Zienkiewicz00,
Rahman91} need excessively large computational resources both in memory and processor usage.\cite{Jiang96,Boriskin08}
This is due to the extremely large computational domain in space and time required to model high quality
modes. To treat such narrow resonances and their perturbations, we
have recently developed\cite{Muljarov10} a rigorous perturbation theory called resonant state
expansion (RSE) and applied it to planar optical resonators with different
perturbations\cite{Doost12} as well as to spherical resonators reducible to effective
one-dimensional (1D) systems.\cite{Muljarov10} We have demonstrated on exactly solvable examples
in 1D that the RSE is a reliable tool for calculation of wave numbers and electro-magnetic fields
of resonant states (RSs),\cite{Doost12} as well as transmission and scattering properties of open
optical systems.

In the present work we report an important step in the development of the RSE,
extending the method to effectively two-dimensional (2D) systems (i.e. 3D systems translational invariant in one
direction) which are {\it not reducible} to effective 1D systems. We
provide a summary of the theory in a general 3D case, apply it to an effective 2D
system, and then consider examples which cover several important types of perturbations. We treat
an ideal dielectric cylinder with uniform dielectric constant in vacuum as unperturbed system and
calculate perturbed RSs for homogeneous (i.e. reducible to 1D) and inhomogeneous perturbations,
including a half-cylinder, thin-film, and thin-wire perturbations. None of these inhomogeneous perturbations
have known exact analytic solutions which could be used for verification of the RSE. However, the
case of a narrow wire inside a cylinder allows for an approximate analytic solution suitable for weak perturbations.\cite{Dettmann08,Dettmann09PRA} This
solution is compared with the present results of the RSE, demonstrating a good agreement.

In the literature, the scattering properties of an open system are often described in terms of a continuum of its eigenstates, all having real frequencies. Such a continuum, together with isolated eigenmodes (e.g. waveguide modes), if they exist in the spectrum, form a mathematically complete set of states suitable for expansions. The continuum, however, presents a significant obstacle for computational methods like perturbation theory, which are based on such expansions. The concept of RSs, naturally following from the observation of resonances in the spectra of open systems, introduces another complete set of eigenstates
which  eliminates the continuum from the spectrum, replacing it by a countable number
of discrete modes with complex frequencies. This is the basis used in the RSE.

One striking feature of 2D systems, which is revealed in the present work,
is the presence of a one-dimensional continuum in the manifold of RSs. This continuum is specific to 2D systems and is required for the completeness of the basis and thus for the accuracy of the RSE applied in 2D as discussed in this
work in some detail. Such continua are generally known in the theory of quasi-guided modes in
photonic crystal structures\cite{Neviere80} as potential sources of Wood-Rayleigh anomalies in
optical spectra.\cite{Wood02,Rayleigh07,Fano41,Hessel65,Akimov11} Technically, they are caused in that case by the presence
of square roots in the photon dispersion of light propagated and Bragg scattered inside the
photonic crystal. In the case of a dielectric cylinder, which is used for the RSE in 2D, the continuum
originates mathematically from the cut in the cylindrical Hankel functions solving Maxwell's equations outside the
cylinder.

The paper is organized as follows. In Sec.II we give the general formulation of the RSE for an
arbitrary three-dimensional (3D) system.  In Sec.\,III we treat the special case of effective 2D systems
using the homogeneous dielectric cylinder as unperturbed system and adding the
contribution of the cut to the RSE. This is followed by examples illustrating the method and
comparing results with existing analytic solutions. Details of the general formulation of the method, its application in 2D,  and the calculation of matrix elements for specific perturbations are given in Appendices A-D.

\section{Resonant state expansion}\label{sec:RSE}
In general, RSs in an open optical system with local dielectric constant $\varepsilon({\bf r})$
and permeability $\mu=1$, where $\mathbf{r}$ is the three-dimensional spatial position, are
the eigensolutions of the Maxwell wave equation,
 \be
 \nabla\times\nabla\times\En(\br)=k_n^2\varepsilon(\br)\En(\br)\,,
 \label{me3D}
\ee which satisfy the outgoing wave boundary condition
 \be
\En(\br)\to r^{-(D-1)/2} e^{i k_n r}\ \ \ {\rm for}\ \  r \to \infty\,,
 \label{BC}
\ee where $r=|\br|$, $D$ is the space dimensionality, $k_n$ is the wave-vector eigenvalue of the
RS numbered by the index $n$, and $\En(\br)$ is its electric field eigenfunction. The
time-dependent part of the RS wave function is given by $\exp(-i\omega_n t)$ with the complex
eigenfrequency $\omega_n=c k_n$, where $c$ is the speed of light in vacuum. RSs are either
stationary or time-decaying solutions of Maxwell's equation. The wave numbers $k_n$ of time-decaying RSs lie in the lower
half of the complex $k$-plane and come in pairs, having the opposite real and equal imaginary parts. Indeed, if  ${\bf E}_n({\bf r})$ and $k_n$ corresponding to RS $n$ satisfy Eqs.\,(\ref{me3D}) and (\ref{BC}),  taking the complex
conjugate of Eq.\,(\ref{me3D}) we find that ${\bf E}(\br)=\En^\ast(\br)$ and $k=\pm k^\ast_n$
also satisfy the same equation. Only $-k^\ast_n$ has a negative imaginary part as required for
time-decaying solutions. We label the resulting RS with index $-n$, so that $k_{-n}=-k^\ast_n$.

For any open system, the RSs form an orthonormal complete set of eigenmodes. It follows from
Eq.\,(\ref{BC}) that solutions decaying in time grow exponentially in space as $r\to \infty$.
Therefore the normalization of RSs cannot be simply given by the usual volume integrals over their
wave functions but it also needs to involve the electromagnetic energy flux through a surface
surrounding the system. The orthogonality of RSs for $n\neq m$ has the form
 \bea
0&=&(k_n^2-k_m^2)\int _V d{\bf r}\varepsilon(\br)\En(\br)\cdot\Em(\br)\nonumber\\
&&-\int _{S_V} dS \left(\En\cdot\frac{\partial\Em}{\partial s}-\Em\cdot\frac{\partial\En}{\partial
s}\right)\,,
 \label{orthog}
 \eea
where the first integral in Eq.\,(\ref{orthog}) is taken over an arbitrary
volume $V$ which include all system inhomogeneities of $\varepsilon({\bf r})$ while the second
integral is taken over the surface $S_V$  surrounding the volume $V$ and contains the
gradients $\partial/\partial s$ normal to this surface. This follows strictly from Maxwell's
equation \Eq{me3D} and Green's theorem, using $\nabla\cdot\En=0$ on the surface $S_V$ which is
situated outside of the system inhomogeneities.
It is convenient, following \Eq{orthog}, to normalize the RSs in the way
 \bea
1&=&\int _V d\br\varepsilon(\br)\En^2(\br)
\nonumber\\
 && -\lim_{k\to k_n}\frac{\displaystyle \int _{S_V} dS
\left(\En\cdot\frac{\partial\mathbf{E}}{\partial s}-\mathbf{E}\cdot\frac{\partial\En}{\partial
s}\right)}{k_n^2-k^2}\,,
 \label{normaliz}
 \eea
where we have used in the second integral an analytic continuation ${\bf E}(k, \br)$ of the RS
wave function $\En(\br)$ around the point $k_n$ in the complex $k$-plane. For solutions decaying
in time both surface and volume integrals in Eqs.\,(\ref{orthog}) and (\ref{normaliz}) diverge for
$V\to \infty$. Their superposition, however, removes the divergencies making the normalization
independent of $V$. In the special case of stationary states which are decaying in space, the
surface integrals vanish for $V\to \infty$ and the usual orthonormality in terms of
infinite-volume integrals is restored.

The RSE is based on the following three key elements. The {\em first} one is the Dyson equation,
\bea \GFPk(\br,\br')&=&\GFk(\br,\br') \label{Dyson}\\
&&-k^2\int \GFk(\br,\br'')\Delta\varepsilon(\br'') \GFPk(\br'',\br') d\br''\,, \nonumber \eea
which relates the perturbed and unperturbed GFs, $\GFPk$ and $\GFk$, respectively.
The difference between the perturbed and unperturbed systems is a perturbation of the dielectric constant $\Delta\varepsilon
(\br)$ with compact support.

 The {\em second} key element is the spectral representation of the GF,
\be \GFk(\br,\br')=\lefteqn{\sum_n}\int \,\frac{\En(\br)\otimes\En(\br')}{k(k-k_n)}\,\frac{k_n}{w_n}\,,
\label{ML4}\ee
which takes into account simple poles of the GF at $k=k_n$ as a sum and a cut of the GF in the complex $k$-plane as an integral.
Details of the derivation of \Eq{ML4} in a general 3D case, using the Mittag-Leffler and reciprocity theorems, are given in Appendix~\ref{App:ML}, accounting for the tensor form of the GF of Maxwell's equation.
The presence of the cut is a specific property of 2D systems having a
continuum of RSs in their spectrum. The nature of the cut and its contribution to the RSE is discussed in more detail in Sec.\,\ref{sec:application2D} and in Appendix~\ref{App:GF}. The spectral representation \Eq{ML4} is used in the Dyson equation (\ref{Dyson}), for both unperturbed and perturbed systems, equating the residues at the perturbed poles of both sides of \Eq{Dyson}. The spectral representation of the unperturbed GF has the normalization constants $w_n=2k_n$ following from the normalization condition \Eq{normaliz}, as shown in Appencies\,\ref{App:ML} and \ref{App:GF}. As for the spectral representation of the perturbed GF $\GFPk$, in which $\En$ and $k_n$ are replaced by $\Epn$ and $\varkappa_\nu$, it is not required by the RSE that the perturbed RSs are normalized in the same way, and thus the corresponding normalization constants can be any.

The {\em third} element is the completeness of RSs, which is discussed in Appendix~\ref{App:ML} and mathematically expressed by the closure relation \Eq{Closure}. It allows to expand the perturbed wave functions into the unperturbed ones:
\be \Epn(\br) = \sqrt{\varkappa_\nu}\lefteqn{\sum_n} \int \,
\frac{c_{n\nu}}{\sqrt{k_n}}
\En(\br)\,.\label{expansion} \ee
The expansion coefficients $c_{n\nu}$ are scaled here by $1/\sqrt{k_n}$, in order that the Dyson equation
reduces to a linear symmetric matrix eigenvalue problem\cite{footnote1}
\be \lefteqn{\sum_m} \int\,
\left(\frac{\delta_{nm}}{k_n}+\frac{V_{nm}}{2\sqrt{k_n k_m}}\right)
c_{m\nu}=\frac{1}{\varkappa_\nu} c_{n\nu}\,, \label{RSE} \ee
in which the matrix elements of the perturbation are defined by
\be V_{nm}=\int\Delta\varepsilon(\br)\, \En(\br)\cdot\Em(\br)\,d \br\,. \label{Vnm} \ee
Equations (\ref{expansion}-{\ref{Vnm}), together with the normalization condition \Eq{normaliz}
used for the unperturbed modes, define the method called the RSE.

\section{Application to 2D systems}\label{sec:application2D}

Let us consider systems in 3D space which are homogeneous in one direction (along the unit vector $\za$ of the $z$-axis),
and thus can be reduced to effective 2D systems, as their wavevector component along $\za$ is
conserved and the solution can be separated into a plane wave $\exp(ik_z z)$  and the
remaining $(x,y)$-problem which we express below in polar coordinates $\brho=(\rho,\varphi)$.
For such a system, the solutions of Maxwell's equations split into two groups with orthogonal
polarizations, called transverse electric (TE) and transverse magnetic (TM), where TE (TM) states
have a electric (magnetic) field orthogonal to $\za$. This nomenclature relates to the theory of
waveguides where the light propagating along $\za$ has the dominant component $k_z$ of the wave vector, and the electric (magnetic) field in TE (TM) modes is thus approximately perpendicular to the wave vector. Although we restrict
our treatment here to the opposite limit of $k_z=0$, we follow these adopted notations. In our case, however, the TE (TM) states have the magnetic (electric) field polarization vector strictly parallel to $\za$ and thus normal to the wave vector of light at large distances.

We treat in this work only the $k_z=0$ TM states for which $\En=\za E_n$ and \Eq{me3D} reduces to
\be \left[\frac{\partial^2}{\partial\rho^2}+\frac{1}{\rho}\frac{\partial}{\partial\rho}
+\frac{1}{\rho^2}\frac{\partial^2}{\partial\varphi^2}
+\varepsilon(\rho,\varphi)k_n^2\right]E_n(\rho,\varphi)=0\,. \label{pdisk} \ee
The states are normalized according to \Eq{normaliz} in which the volume $V$ is given by an infinitely long cylinder of radius $R$. In order to make the normalization constants finite, RSs are normalized per unit length along $\za$. The normalization following from \Eq{normaliz} is given more explicitly by\cite{Muljarov10}
\bea &&\!\!\!\!\!\!\!1=\int_0^{2\pi}d\varphi\int_0^R\rho d\rho\,\varepsilon(\rho,\varphi)E_n^2
(\rho,\varphi) \label{normaliz2} \\
 &&\!\!\!\!\!\!\! +\frac{1}{2k_n^2}\int_0^{2\pi}d\varphi \left[E_n\frac{\partial {
E}_n}{\partial \rho}+\rho E_n\frac{\partial^2 E_n}{\partial \rho^2}-\rho\left(\frac{\partial
E_n}{\partial \rho}\right)^2\right]_{\rho=R_+},\nonumber \eea
where the last integral is taken over the outer surface of the cylinder. We choose as
unperturbed system for the RSE a homogeneous dielectric cylinder in vacuum with radius $R$ and refractive index $n_r$ (which is neither 0 nor 1), having
\be \varepsilon(\rho,\varphi)=\left\{
\begin{array}{lll}
n_r^2 & {\rm for} & \rho\leqslant R\,, \\
1 & {\rm for} & \rho>R\,.
\end{array}
\right.\label{epsilon} \ee
Due to the cylindrical symmetry, the azimuthal index $m$ is a good quantum number which takes
integer values giving the number of field oscillations around the cylinder. The unperturbed RS
wave functions factorize as
\be E_n (\rho,\varphi)=R_m(\rho,k_n) \chi_m(\varphi)\,, \label{E-fact} \ee
where the angular parts are defined by
\be \chi_m(\varphi)=\left\{
\begin{array}{lll}
\pi^{-1/2}\sin(m\varphi) & {\rm if} & m<0\,, \\
(2\pi)^{-1/2} & {\rm if} & m=0\,, \\
\pi^{-1/2}\cos(m\varphi) & {\rm if} & m>0\,,
\end{array}
\right. \label{chi-n} \ee
and are orthonormal according to
\be \int_0^{2\pi} \chi_m(\varphi)
\chi_{m'}(\varphi)d\varphi =\delta_{mm'}\,. \ee
The choice of the wave functions in the form of standing waves \Eq{chi-n}, instead of the more usual $e^{im\varphi}$, is dictated by the general orthogonality condition defined by \Eq{orthog}, without using the complex conjugate. The radial components have the form
\be R_m(\rho,k)= A\left\{
\begin{array}{lll}
J_m(n_r k\rho)/J_m(n_r kR) & {\rm for} & \rho\leqslant R\,, \\
H_m(k\rho)/H_m(kR) & {\rm for} & \rho>R\,, \\
\end{array}
\label{R-analyt} \right. \ee
in which $J_m(z)$ and $H_m(z)\equiv H_m^{(1)}(z)$ are, respectively, the cylindrical Bessel and
Hankel functions of the first kind. The wave functions are normalized according to \Eq{normaliz2}
with the normalization constant
\be A=\frac{1}{R}\sqrt{\frac{2}{n_r^2-1}}\,.
\label{A-norm}\ee
The two boundary conditions at the surface of the cylinder, the continuity of the electric field
and its radial derivative, produce a secular equation for the RS wave number eigenvalues $k_n$,
which has the form
\be D_m(k_nR)=0\,,\label{secular2} \ee
where
\be D_m(z)=n_rJ_m'(n_rz)H_m(z)-J_m(n_rz)H_m'(z)\,, \label{Dfunc} \ee
and $J_m'(z)$ and $H_m'(z)$ are the derivatives of $J_m(z)$ and $H_m(z)$, respectively. Here $z$ represents a complex argument, as opposed to the spatial coordinate used earlier.

The Hankel function $H_m(z)$ which describes the field outside the cylinder and
contributes to Eqs.\,(\ref{R-analyt}) and (\ref{Dfunc}) is a multiple-valued function, or in other words is defined on a
Riemann surface having infinite number of sheets due to its logarithmic component. However, only one of these sheets contains the eigenvalues $k_n$, satisfying \Eq{secular2}, which correspond to the outgoing wave boundary conditions of RSs.
This `physical' sheet of $H_m(z)$ has a cut in the complex $z$-plane along the negative imaginary half-axis, as shown in Appendix~\ref{App:GF}, which in turn gives rise to the same cut in the GF.
Consequently, the Mittag-Leffler theorem used for the spectral representation of the GF needs to be modified
to include the cut contribution, as done in Appendix\,\ref{App:GF}.

For the TM case treated here, the full GF of the homogeneous dielectric cylinder, which is defined via Maxwell's equation with a line current source term, is given by $\GFk=G_k\,\za\otimes \za$,
in which
\be G_k(\brho,\brho')=\sum_m G_m(\rho,\rho';k) \chi_m(\varphi)\chi_m(\varphi')\,,
\label{GF-TM} \ee
and the radial components have the following spectral representation
\bea
G_m(\rho,\rho';k)&=&\sum_n\frac{R_m(\rho,k_n)R_m(\rho',k_n)}{2k(k-k_n)}\nonumber\\
&&+\int_{-i\infty}^0 \frac{R_m(\rho,k')R_m(\rho',k')}{2k(k-k')}\sigma_m(k') dk'\nonumber\\
&\equiv&\lefteqn{\sum_n} \int \,\frac{R_m(\rho,k_n)R_m(\rho',k_n)}{2k(k-k_n)} \label{Gm} \eea
derived in Appendix~\ref{App:GF}. Note that the cut contribution to the GF spectrum in the form of
the integral in the last equation is described in terms of the {\em same functions} as those used
for discrete poles. This implies that the cut of the GF can be understood as a continuous distribution of
additional poles along the negative imaginary half-axis with the density
\be \sigma_m(k)=\frac{4(n_r^2-1)J_m(n_rkR)}{\pi^2 k D_m^+(kR) D_m^-(kR)} \label{sigma} \ee
calculated in Appendix~\ref{App:GF}. Here $D^{\pm}_m(z)$ are the two limiting values of $D_m(z')$ for
$z'$ approaching point $z$ on the cut from its different sides Re$\,z'\gtrless 0$. Remarkably, the integrated
density of the cut contribution to the GF is equivalent to half a normal pole:
$\int_{-i\infty}^0 \sigma_m(k) d k = (-1)^{m+1}/2$.

\begin{figure}[t]
\includegraphics*[width=0.95\columnwidth]{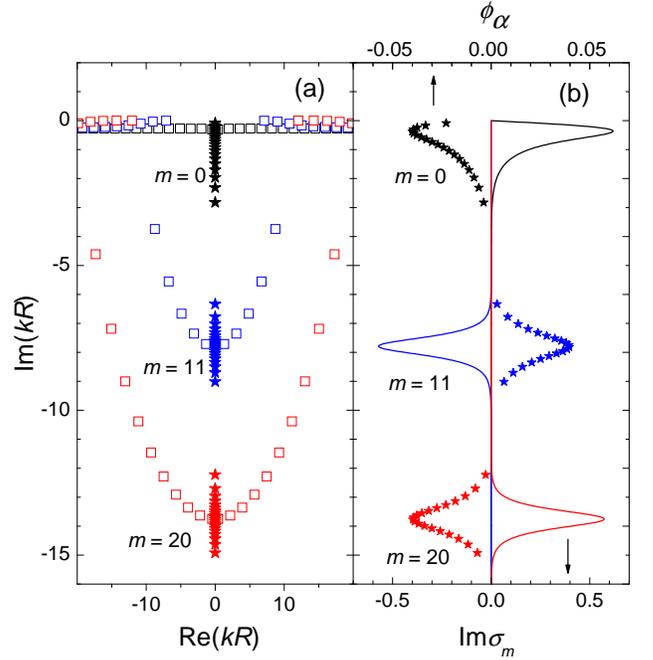}
\caption{(a): Cut poles $k_\alpha$ (stars)  representing the cut of the GF of a homogeneous dielectric cylinder
with ${n_r}=2$, in the complex wave-number plane for $m=0$, 11, and 20. Normal poles
$k_n$ (open squares) are also shown. (b): Cut pole density $\sigma_m(k)$ (solid curves) and the cut pole strength $\phi_\alpha^{(m)}$ (stars), for the same values of $m$.
}\label{fig:Fict}
\end{figure}

To numerically treat the cut contribution in the linear eigenvalue problem \Eq{RSE}, we
discretize the integral in \Eq{Gm} into a finite number of cut poles and add cut RSs to the basis.
These cut poles have non-integer strength $\phi_\alpha^{(m)}$ determined by the cut pole
density $\sigma_m$. The function $\sigma_m(k)$ is purely imaginary and is peaked close to normal
poles $k_n$ as can be seen in \Fig{fig:Fict} for selected $m$. In the numerical calculations of
the present work we have used cut pole positions and strengths determined by splitting the cut
interval $[0,-i\infty]$ into $N_c^{(m)}$ regions $[q_\alpha^{(m)},q_{\alpha+1}^{(m)}]$
numbered by $\alpha=1\,,2\,,\dots,N_c^{(m)}$, which are chosen to contain an equal weight according to
\be \int_{q_\alpha^{(m)}}^{q_{\alpha+1}^{(m)}}\sqrt{|\sigma_m(k)|}dk = \frac{1}{N_c^{(m)}}
\int_{-i\infty}^{0}\sqrt{|\sigma_m(k)|}dk. \label{weight} \ee
For the numerical results shown later in this section and for the chosen values of $N_c^{(m)}$,
using the weight $\sqrt{|\sigma_m|}$ in \Eq{weight} was found to give the best accuracy of the RSE as compared to
other powers of $|\sigma_m|$.
Each region $[q_\alpha^{(m)},q_{\alpha+1}^{(m)}]$ of the cut is represented by a cut pole of the GF at $k=k_\alpha^{(m)}$
given by the first moment,
\bea k_\alpha^{(m)}=\left.\int_{q_\alpha^{(m)}}^{q_{\alpha+1}^{(m)}}k\sigma_m(k)dk\right/\phi_\alpha^{(m)}\,,
\label{k-alpha} \eea
where the cut pole strength $\phi_\alpha^{(m)}$ is defined as
\be
\phi_\alpha^{(m)}=\int_{q_\alpha^{(m)}}^{q_{\alpha+1}^{(m)}}\sigma_m(k)dk\,.
\label{phi-m-alpha}
\ee

An example of cut poles assigned for $m=0$, 11, and 20 is given in \Fig{fig:Fict}(a).
The cut poles contribute to the RSE in the same way as the normal poles, and the matrix elements with the cut RSs
are given by the overlap integrals \Eq{ME-cut} expressed in terms of exactly the same functions \Eq{R-analyt} as for the normal RSs. In discretization of the linear eigenvalue problem \Eq{RSE} of the RSE, the only modification caused by the cut is that the matrix of the perturbation is weighted according to the cut pole strengths $\phi^{(m)}_\alpha$, as described by \Eq{RSE2} in Appendix~\ref{App:RSEcut}.

In the numerical calculation, the total number of poles $N_t$ used in \Eq{RSE} determines
the computational complexity of the matrix eigenvalue problem, so that we are interested in the number of cut poles in the basis
producing the best accuracy for a given $N_t$. We have investigated this numerically for the examples given below,
and found that this is achieved using about 20\% cut poles in the basis. Only for the homogeneous
perturbation in Sec.\,\ref{sec:Hom}, we used $N_c^{(m)}\sim N^{(m)}$, where $N^{(m)}$ is the number of normal poles in the basis  for the given $m$, in order to demonstrate the convergence towards the
exact solution. For all other numerical results we used $N_c^{(m)}\sim0.2N^{(m)}$.

\bigskip

The rest of this section discusses results of the RSE for different effective 2D systems. We
consider a homogeneous dielectric cylinder of radius $R$ and refractive index $n_r=2$
($\varepsilon=4$) with different types of perturbations, namely a homogeneous perturbation of the
whole cylinder in \Sec{sec:Hom}, a half-cylinder perturbation in \Sec{sec:Halfmoon}, a thin-film
perturbation in \Sec{sec:Line} and a wire perturbation in \Sec{sec:Delta}. Explicit forms of the
matrix elements for these perturbations and details of their calculation are given in
Appendix~\ref{App:ME}.
\begin{figure}[b]
\includegraphics*[width=0.95\columnwidth]{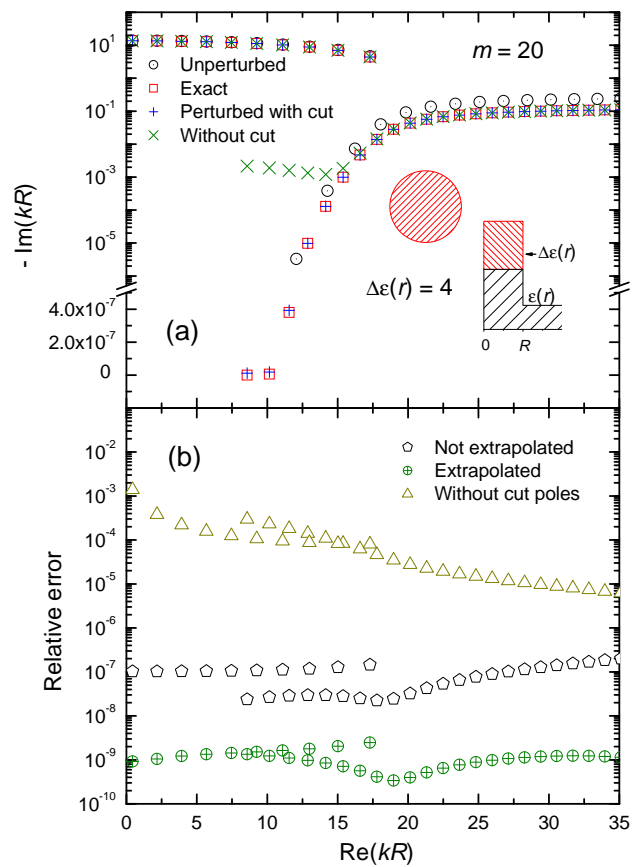}
\caption{(a): Perturbed RS wave numbers for the homogeneous perturbation Eq.\,(\ref{eps-hom})
calculated via the RSE with $N=800$ (only sine modes are shown). The perturbed poles with (+) and without ($\times$) the cut
contribution are compared with the exact solution (open squares). Unperturbed wave numbers are
also show (open circles with dots). Inset: Dielectric constant profile for the unperturbed and perturbed
systems. (b): Relative error in the calculated
perturbed wave numbers with (heptagons) and without (triangles) contribution of the cut. Relative
error for a simulation including the cut and improved by extrapolation is also shown (crossed
circles). }\label{fig:F2}
\end{figure}

\subsection{Homogeneous Cylinder Perturbation}
\label{sec:Hom} The perturbation we consider in this section is a homogeneous change of
$\varepsilon$ over the whole cylinder, given by
\be \Delta\varepsilon(\rho,\varphi)=\Delta\varepsilon \theta(R-\rho)=\left\{
\begin{array}{cl}
\Delta\varepsilon & \text{for\ \  } \rho \leqslant R\,,\\
0 &\text{for\ \  } \rho > R
\end{array} \right.
\label{eps-hom} \ee
with the strength $\Delta\varepsilon=4$ used in the numerical calculation. For
$\varphi$-independent perturbations, modes with different azimuthal number $m$ are decoupled, and so
are even and odd (cosine and sine) modes given by \Eq{chi-n}. We show only the sine modes here, and use
for illustration $m=20$. The matrix elements of the perturbation are calculated
analytically and given by Eqs.\,(\ref{V-homogeneous1}) and (\ref{V-homogeneous2}). The homogeneous
perturbation does not change the symmetry of the system, so that the perturbed modes obey the same
secular equation \Eq{secular2} with the refractive index $n_r$ of the cylinder changed to
$\sqrt{n_r^2+\Delta\varepsilon}$, and thus the perturbed wave numbers $\varkappa_\nu$ calculated
using the RSE can be compared with the exact values $\varkappa^{\rm (exact)}_\nu$.

We choose the basis of RSs for the RSE in such a way that for the given azimuthal number $m$ and the given number of normal RSs $N$ we find all normal poles $|k_n|<k_{\rm max}(N)$ with a suitably chosen maximum wave vector $k_{\rm max}(N)$ and then add the cut poles.
We find that as we increase $N$, the relative error $\bigl|{\varkappa_\nu}/{\varkappa^{\rm
(exact)}_\nu}-1\bigr|$ decreases as $N^{-3}$. Following the procedure described in
Ref.\,\onlinecite{Doost12} we can extrapolate the perturbed wave numbers. The resulting
perturbed wave numbers are shown in \Fig{fig:F2}. The perturbation is strong, creating 3 additional
WGMs with $m=20$ having up to 4 orders of magnitude narrower linewidths. For $N=800$, the RSE reproduces about 100 modes to a relative error in the $10^{-7}$ range, which is decreasing by one or two orders of magnitude after extrapolation. The contribution of the cut is
significant: Ignoring the cut leads to a relative error of the poles in the $10^{-3}$ range. The fact that the relative error improves by 4-5 orders of magnitude after taking into account the cut in the form of the cut poles
shows the validity of the reported analytical treatment of cuts in the RSE, and the
high accuracy of the discretization method into cut poles.

\subsection{Half-Cylinder Perturbation}
\label{sec:Halfmoon} We now consider a bulk perturbation which mixes modes with different $m$. The
perturbation is given by
\be \Delta\varepsilon(\rho,\varphi)=\Delta\epsilon\, \theta(R-\rho)\times
\left\{
\begin{array}{rl}
1 & \text{for\ \  } |\varphi| \leqslant {\pi/2}\,,\\
-1 &\text{otherwise. }
\end{array} \right.
\label{halfmoon} \ee In our numerical simulation we take $\Delta\epsilon=0.2$. The matrix elements
of the perturbation are given by Eqs.\,(\ref{V-halfmoon1})--(\ref{Qmm}) which require a numerical
integration. Owing to the symmetry of the perturbation, the sine and cosine basis modes are still
de-coupled, therefore we treat them separately, see panels (a) and (b) in Fig.\,\ref{fig:F3}. Due
to a relatively small perturbation (compared to that considered in Sec.\,\ref{sec:Hom}), the mode
positions in the spectrum do not change much. However, the quality factors $Q$ of all WGMs
decrease, as the lifetime of the resonances is now limited by an additional scattering at the step
in the dielectric constant of the perturbed cylinder.

\begin{figure}[b]
\includegraphics*[width=0.95\columnwidth]{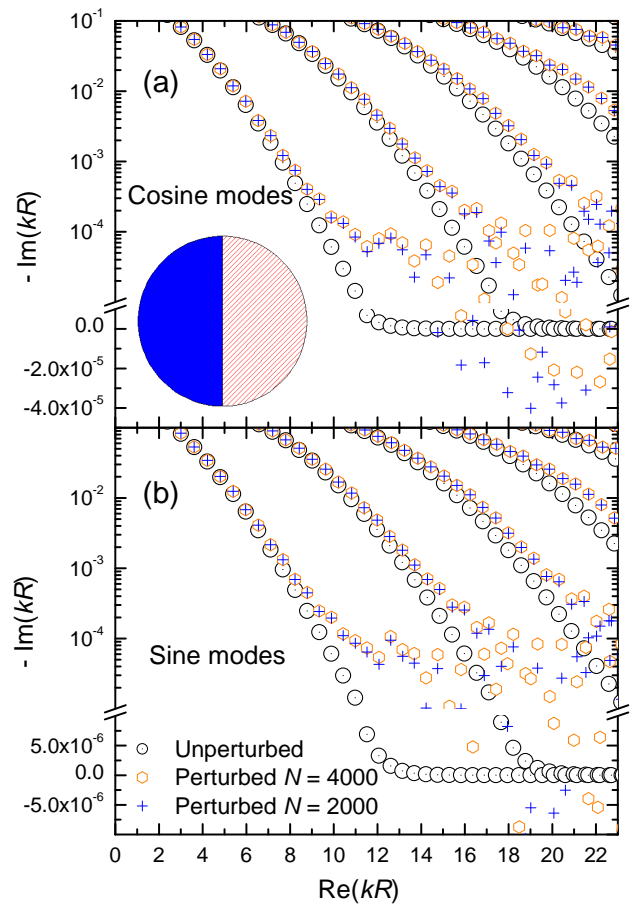}
\caption{Unperturbed (open circles with dots) and perturbed RS wave numbers of (a) cosine and (b) sine modes
of a cylinder for a half-cylinder perturbation Eq.\,(\ref{halfmoon}) with $\Delta\varepsilon=0.2$
and the basis sizes $N=2000$ (crosses) and $N=4000$ (hexagons). Only the WGM region is shown. Inset:
Diagram showing the regions of increased (solid blue) and decreased (red and white striped)
dielectric constant. }\label{fig:F3}
\end{figure}
\begin{figure}
\includegraphics*[width=0.95\columnwidth]{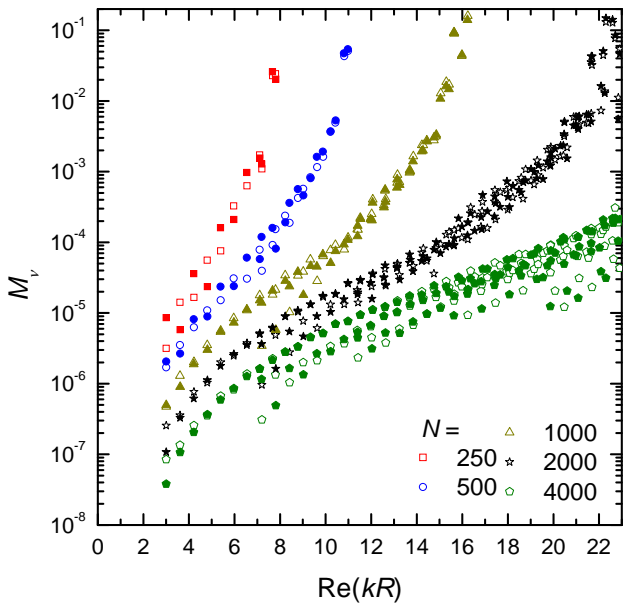}
\caption{Absolute errors $M_{\nu}$ of the RS wave numbers $\varkappa_\nu$ for the half-cylinder perturbation Eq.\,(\ref{halfmoon}) as functions of ${\rm Re}\,\varkappa_{\nu}$, calculated via the RSE for different basis sizes $N$, for cosine (closed
shapes) and sine (open shapes) modes.}\label{fig:F4}
\end{figure}%
To the best of our knowledge, an analytic solution for this perturbation is not available and thus we cannot calculate
the relative error of the RSE result with respect to the exact solution. However, we can investigate the
convergence of the method in order demonstrate how the RSE works in this case which is not
reducible to an effective one-dimensional problem. This is done in \Fig{fig:F3} showing the
perturbed sine and cosine modes for two different values of basis size $N$ and in \Fig{fig:F4}
where absolute errors $M_{\nu}$ are shown for several different values of $N$. Following
Ref.\,\onlinecite{Doost12}, the absolute error is defined here as
$M_{\nu}=\max_{i=1,2,3}|\varkappa_\nu^{N_4}-\varkappa_\nu^{N_i}|$, where $\varkappa_\nu^{N_i}$ are
the RS wave numbers calculated for basis sizes of $N_1\approx N/2$, $N_2\approx N/\sqrt{2}$,
$N_3\approx N/\sqrt[4]{2}$, and $N_4=N$. The results for the cosine and sine modes are quite
similar. From \Fig{fig:F4} we see that the perturbed resonances are converging with increasing basis size. Though, the absolute error has some fluctuations within an order.

We were able to see the power law of the convergency, in agreement with \Onlinecite{Doost12}, and
found that the power law exponent is approximately $-2$. We found however that owing to the above
mentioned fluctuations, the power law convergence is not well developed compared to the
one-dimensional problems considered so far (including the example of \Sec{sec:Hom}). Increasing
the basis size $N$ improves the power-law convergency. This is attributed to the larger number of
basis states below a given $|k_{\rm max}R|$ in effective 2D systems compared to effective 1D systems. Thus one needs a larger basis in order to approximate discrete steps of the basis size by a continuous power law.

To show how a particular perturbed state is created as superposition of unperturbed states, we
show by a star in \Fig{fig:COSN4000} one the perturbed WGMs of \Fig{fig:F3}\,(a) which, owing to the perturbation,
increases its linewidth by nearly an order of magnitude. The contribution of the basis states is visualized by circles of a
radius proportional to $|c_{n\nu}|^{1/3}$, which are centered at the positions of the wave vectors $k_n$ in
the complex $k$-plane. The expansion coefficients $c_{n\nu}$ decrease quickly with the distance to the spectral
position of the perturbed mode $\varkappa_\nu$, with the dominant contribution coming from the nearest unperturbed
RS, a typical feature of perturbation theory in closed systems. Importantly, this demonstrates
that if we are interested in the  modes within a small spectral region, we can limit the basis in the
RSE to states close to that region. This result is crucial for the application of the RSE to
effective 3D systems which have even larger numbers of basis states below a given $|k_{\rm max}R|$, as one
can significantly reduce the number of basis states needed to calculate the perturbation of a mode
of interest to a given accuracy.

\begin{figure}[t]
\includegraphics*[width=0.95\columnwidth]{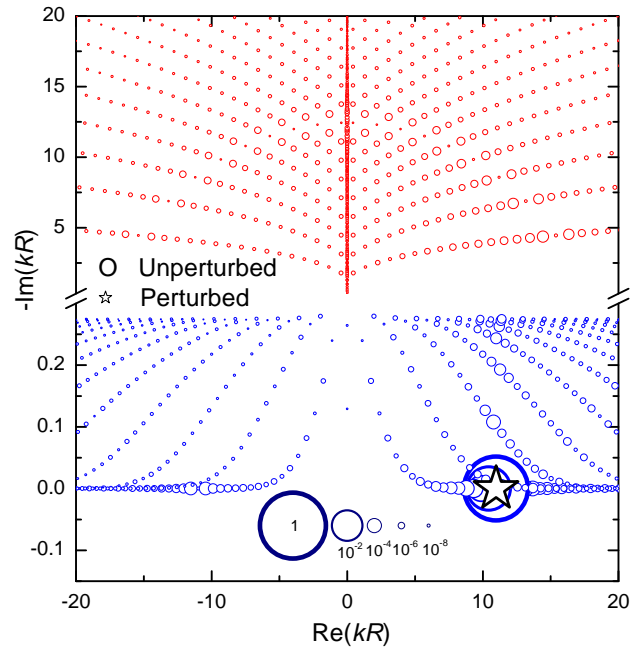}
\caption{Contributions of the basis RSs (black circles) to a given perturbed RS (blue star), calculated for
the cosine modes and the half-cylinder perturbation \Eq{halfmoon}, using $N=4000$. All circles and
the star are centered at the positions of the corresponding RS wave numbers in the complex
$k$-plane.  The radius of the black circles is
proportional to $|c_{n\nu} |^{1/3}$. A key showing the relationship between circle radius and
$|c_{n\nu}|^{2}$ is given by the red circles.}\label{fig:COSN4000}
\end{figure}

\subsection{Thin-Film Perturbation}
\label{sec:Line}
We now move from the bulk perturbations towards the case of a thin film embedded in the cylinder,
corresponds to a line perturbation in the effective 2D system. The perturbation we consider in
this section is given by
\be \Delta\varepsilon(\rho, \varphi)=h\Delta\varepsilon
\,\frac{\theta(R-\rho)}{\rho}\,\delta(\varphi)\,, \label{line} \ee
see the inset in Fig.\,\ref{fig:F5}. In our numerical simulation we take the strength of the
perturbation $h\Delta\varepsilon=-0.1R$. Physically this perturbation
corresponds to a thin metal film of uniform negative dielectric constant $n_r^2+\Delta\epsilon$
and width $h$ much narrower than the shortest wavelength in the basis. The perturbation leaves the
sine modes of the unperturbed cylinder unchanged. Hence we only include cosine modes into the
basis. The perturbation matrix elements are given by \Eq{V-line}.

\begin{figure}
\includegraphics*[width=0.95\columnwidth]{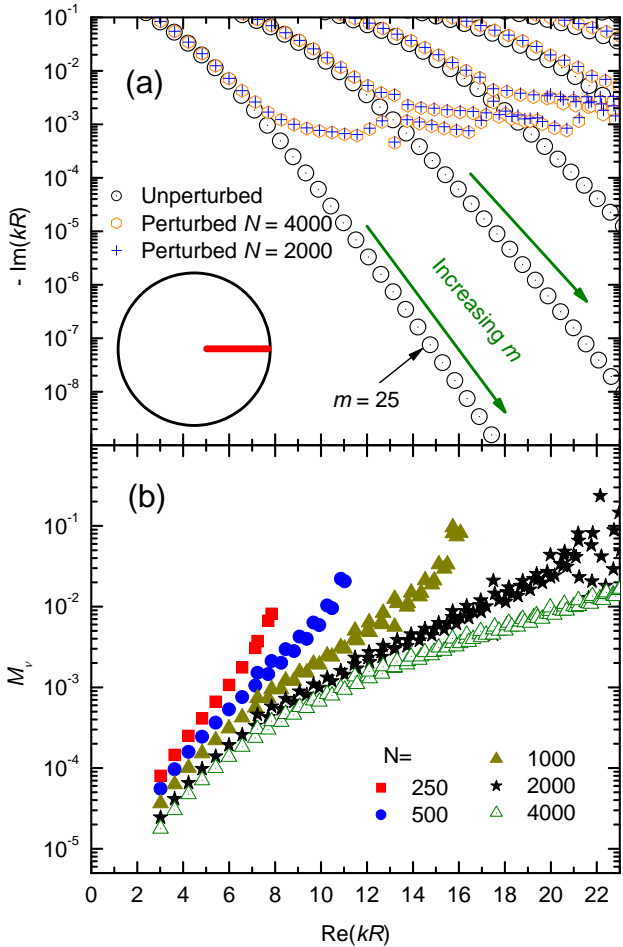}
\caption{(a): Unperturbed and perturbed RS wave numbers of cosine modes for a thin-film
perturbation given by \Eq{line} with $h\Delta\varepsilon=-0.1R$, calculated via the RSE with the
basis sizes $N=2000$ (crosses) and $N=4000$ (hexagons). The unperturbed RSs are shown as open
circles with dots. (b): Absolute errors $M_{\nu}$ as functions of ${\rm Re}\,\varkappa_{\nu}$ calculated for
different basis sizes $N$ as labeled. Inset: Sketch showing the location of the thin metal film
perturbation as a red line inside the unperturbed cylinder. }\label{fig:F5}
\end{figure}

\begin{figure}[t]
\includegraphics*[width=0.95\columnwidth]{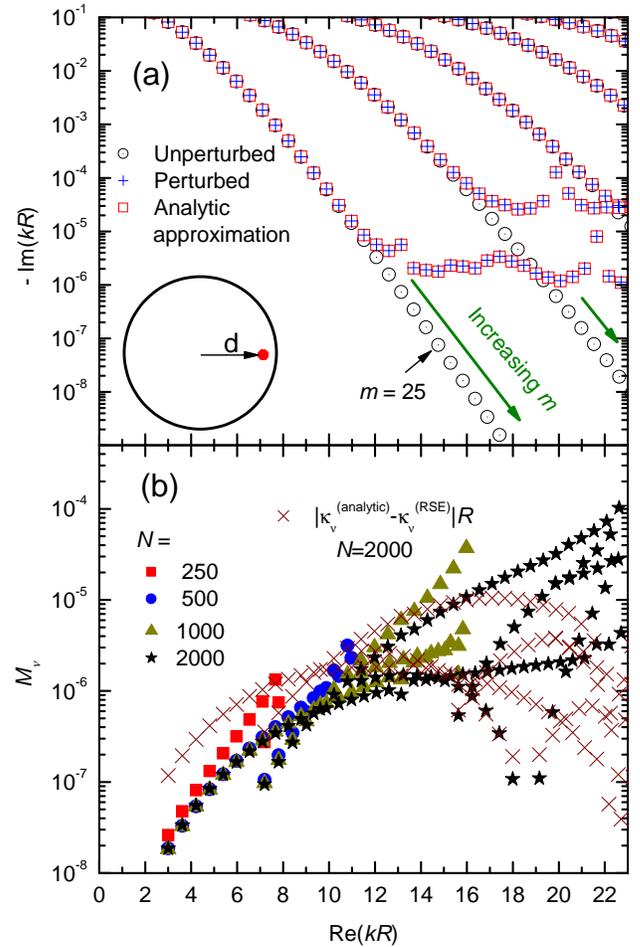}
\caption{(a): Unperturbed and perturbed RS wave numbers of cosine modes for a thin-wire
perturbation given by \Eq{point} with $d=0.8R$, $b=0.001R$, and $\Delta\varepsilon=100$,
calculated via the RSE with the basis size $N=2000$ (crosses) and compared with the  analytic
approximation of Ref.\,\onlinecite{Dettmann09PRA} (empty squares). The unperturbed basis states are
shown as open circles with dots. (b): Absolute errors $M_{\nu}$ in RSE as functions of ${\rm Re}\,\varkappa_{\nu}$
calculated for different basis size $N$ and the absolute difference between the RSE and the analytic approximation (crosses $\times$). Inset: Sketch showing the location of the wire as a
red dot inside the unperturbed cylinder.} \label{fig:F6}
\end{figure}

To our knowledge an analytic solution for this perturbation is not known and we therefore
calculate the absolute error as in \Sec{sec:Halfmoon}. \Fig{fig:F5} shows the resulting RS wave
numbers and absolute errors for this thin-film perturbation. We see in \Fig{fig:F5}(b) that the
convergence of the RSE is slower than in the case of the half-moon perturbation. This is expected
as the thin film has no geometrical effect on the wave-vector $k_y$, giving higher contributions
of basis states with large $k_y$, similar to the results in 1D with a delta scatterer perturbation
reported earlier.\cite{Doost12}
We have found that the power law exponent in this case is approximately $-1$.

\subsection{Thin-Wire Perturbation}
\label{sec:Delta} As last example we consider a dielectric cylinder perturbed by a thin-wire
perturbation which is represented by small disk of radius $b$ centered at the point ${\bf d}$ on
the $x$-axis $(\varphi=0)$. We do not use here a delta perturbation, in order to compare it with an
analytic solution available in the literature.\cite{Dettmann09PRA} The perturbation is defined as
\be \Delta\varepsilon(\brho)=\Delta\varepsilon\,\theta (b-|\brho-{\bf d}|)\,, \label{point} \ee
and we choose $d=|\mathbf{d}|=0.8R$, $b=0.001R$, and $\Delta\epsilon=100$ (the
unperturbed system is the same as before having $n_r=2.0$). This perturbation leaves the sine and
cosine modes decoupled, and the sine modes approximately unchanged (strictly for $d\to0$).
Therefore we show here the perturbation of the cosine modes. The RSE perturbation matrix elements
are given by \Eq{V-delta}. The resulting RS wave numbers are shown in \Fig{fig:F6}\,(a) together
with the analytic approximation, demonstrating a good agreement.

The absolute errors $M_{\nu}$ are shown in \Fig{fig:F6}\,(b). We see that
the convergence in the case of a thin wire is even slower than for the thin-film perturbation
shown in \Sec{sec:Line}. This is expected as the thin wire has no geometrical effect on both
$k_x$ and $k_y$, giving higher contributions of the basis states with large
$|k|$. We found that within the basis sizes investigated, the power law is not well developed, but
for weaker (smaller $|\Delta\epsilon|$) or more spatially extended (larger $b$) perturbations a
better convergence is observed, as expected.

The analytic solution of \Onlinecite{Dettmann09PRA} for a point-like scatterer in a 2D disk is not
strict in any physical system. In the case of a delta scatterer, the secular equation is
logarithmically divergent and thus cannot be used, while the accuracy of the model for a finite
size scatterer relies on a number of approximations\cite{Dettmann08,Dettmann09PRA} which require
$|n_r \varkappa_{\nu}b|\ll1$, $|\varkappa_{\nu}b \sqrt{n_r^2+\Delta\epsilon}|\ll1$, and also
$|{\rm Re}\, \varkappa_\nu| \gg|{\rm Im}\, \varkappa_\nu|$, i.e. having a large $Q$. In addition to this, the
point-like perturbation should not be too close to the edge of the disk, i.e. $|n_r\varkappa_\nu(R-d)|\ll 1$. While we have chosen our parameters to be suitable for these approximations, we do not have a quantitative estimate of the error.
Nevertheless, the comparison in
\Fig{fig:F6}\,(a) of the RSE calculation with the analytic solution demonstrates a good agreement
which is improving as we move closer to the origin in the complex $k$-plane, as detailed in \Fig{fig:F6}\,(b) where the absolute difference between the two calculation is shown.

\section{Summary}\label{sec:summary}
We have applied the resonant state expansion (RSE) to effective two-dimensional (2D) open optical
systems, such as dielectric micro-cylinders and micro-disks with perturbations. We have found and
treated a cut of the Green's functions (GFs) of effective 2D systems -- a feature which to our knowledge has not been
mentioned in the literature but turned out to be crucial for the RSE as the states on the cut
contribute to the completeness of the basis of RSs needed for the accuracy of the method. We have
detailed the formulation of the RSE for a general 3D case taking into account the vectorial nature
of the electro-magnetic field and tensor form of the GF and shown in detail how the theory is
applied to effective 2D systems for which states on the cut are introduced and discretized for the
numerics.

Using the analytically known basis of resonant states (RSs) of an ideal homogeneous dielectric
cylinder -- a complete set of eigenmodes satisfying outgoing wave boundary conditions -- we have
treated different types of perturbations, such as half-cylinder, thin-film and thin-wire
perturbations. For all of these perturbations, the perturbed systems are not reducible to
effective 1D ones, so that the present work demonstrates the applicability of the RSE to general
effective 2D perturbations which mix all basis modes. We investigated the convergency for these
perturbations and compared the RSE results, where it was possible, with available analytic
solutions. In particular, we have made such a comparison for a homogeneous perturbation of a
cylinder, which is reducible to an effective 1D system, and for point-like perturbation of a disk
which presents an essentially 2D system with mixing of all kind of modes in the given polarization
of light. In both situation we have found agreement between the RSE and the known analytic
solutions.

\acknowledgments M.\,D. acknowledges support by EPSRC under the DTA scheme.
\appendix
\section{Spectral representation of the Green's function of an open system}
\label{App:ML}

The Green's function (GF) of an open electromagnetic system is a tensor $\GFk$ which satisfies the outgoing wave
boundary conditions and the Maxwell wave equation (\ref{me3D}) with a delta function source term,
 \be
- \nabla\times\nabla\times \GFk(\br,\br')+k^2\varepsilon({\bf r})\GFk(\br,\br')=\hat{
\mathbf{1}}\delta(\br-\br')\,,
 \label{GFequ}
\ee where $\hat{\mathbf{1}}$ is the unit tensor and $k=\omega/c$ is the wave vector of the
electro-magnetic field in vacuum determined by the frequency $\omega$, which can be real or
complex. Physically, the GF describes the response of the system to a point current with frequency
$\omega$, i.e. an oscillating dipole. Using the reciprocity theorem,\cite{Born99} the relation
 \be
{\bf d}_1\GFk(\br_1,\br_2) {\bf d}_2={\bf d}_2\GFk(\br_2,\br_1) {\bf d}_1
 \label{reciprocity}
\ee holds for any two dipoles ${\bf d}_{1,2}$ at points $\br_{1,2}$ oscillating with the same
frequency. Therefore $\GFk(\br,\br')$ is a symmetric tensor. Assuming  a simple-pole structure of the GF with poles at $k=q_n$ and
taking into account its large-$k$ vanishing asymptotics, the Mittag-Leffler
theorem\cite{More71,Bang81} allows us to express the GF as \be
\GFk(\br,\br')=\sum_n\frac{\Qn(\br,\br')}{k-q_n}\,.
 \label{ML1}
\ee Substituting this expression into \Eq{GFequ} and convoluting with an arbitrary finite field
${\bf D}({\bf r})$ over a finite volume $V$ we obtain
$$
\sum_n \frac{- \nabla\times\nabla\times\Fn({\bf r})+k^2\varepsilon(\br)\Fn(\br)}{k-q_n}={\bf
D}(\br)\,,
$$
where $\Fn(\br)=\int_V \Qn(\br,\br'){\bf D}(\br')d\br'$. Taking the limit $k\to q_n$ yields
$$
- \nabla\times\nabla\times\Fn(\br)+q_n^2\varepsilon(\br)\Fn(\br)=0\,.
$$
Due to the convolution with the GF, $\Fn(\br)$ satisfies the same outgoing wave boundary conditions Eq.\,(\ref{BC}). Then, according to \Eq{me3D},   $\Fn(\br)\propto \En(\br)$ and  $q_n=k_n$. Note that the convolution of the kernel $\Qn(\br,\br')$ with different functions ${\bf D}(\br)$ can be proportional to one and the same function $\En(\br)$ only if the kernel has the form of a direct product:
\be \Qn(\br,\br')=\En(\br)\otimes\En(\br')/w_n\,, \label{QnE}\ee
in which the vectorial nature of the field and the symmetry of the kernel \Eq{reciprocity} were taken into account and
a normalization constant $w_n$ was introduced.
The direct vector product $\otimes$ is defined as ${\bf c}({\bf a}\otimes{\bf b}){\bf d}=({\bf
c}\cdot{\bf a})({\bf b}\cdot{\bf d})$, for any vectors ${\bf a}$, ${\bf b}$, ${\bf c}$, and ${\bf
d}$.

The asymptotics of the GF for $k\to \infty$ following from \Eq{GFequ} requires
that $\GFk\propto k^{-2}$, which for the spectral representation \Eq{ML1} provides the sum
rule:\cite{Bang81}
\be \sum_n \Qn(\br,\br')=0\,. \label{Sum-rule} \ee
Using \Eq{QnE} and \Eq{Sum-rule}, together with \Eq{ML1} yields
\be \GFk(\br,\br')=\sum_n \frac{\En(\br)\otimes\En(\br')}{k(k-k_n)}\,\frac{k_n}{w_n}\,.
\label{ML2} \ee
The normalization constants $w_n$ can be determined from the normalization condition
\Eq{normaliz}. We have shown for specific systems with analytic solutions\cite{Muljarov10} that
\be w_n=2k_n\,. \label{wn} \ee
These systems were 1D or 3D with spherical symmetry, where the normalization constants $w_n$ have
the meaning of the Wronskian derivatives at $k=k_n$. The derivation for 2D systems with
cylindrical symmetry is similar. However a specific property of 2D systems is the presence of a
continuum in the RS spectrum, which is a cut of the GF in the complex $k$-plane. The spectral
representation of the GF includes this cut contribution as integral:
\be \GFk(\br,\br')=\lefteqn{\sum_n}\int \,\frac{\En(\br)\otimes\En(\br')}{2 k(k-k_n)}\,.
\label{ML3}\ee
The cut is discussed in more detail in Sec.\,\ref{sec:application2D} and in Appendix~\ref{App:GF}, where we also show the validity of \Eq{wn} for a homogeneous cylinder.

Substituting the GF of \Eq{ML3} into \Eq{GFequ} and using the sum rule \Eq{Sum-rule} leads to the closure
relation
\be \frac{\varepsilon(\br)}{2} \lefteqn{\sum_n}\int \,\En(\br)\otimes\En(\br')=\hat{\mathbf{
1}}\delta(\br-\br') \label{Closure} \ee
which expresses the completeness of the RSs, so that any function can be written
as a superposition of RSs including the contribution of the cut. Due to the surface term in the orthogonality condition \Eq{orthog}, RSs
form an over-complete basis,\cite{Garcia82,Lind93} i.e. any RS wave function $\En(\br)$ can be
written as a superposition of the wave functions of other RSs of the basis.

\section{Green's function of a homogeneous cylinder}
\label{App:GF}

The TM component of the GF of a homogeneous cylinder in vacuum satisfies the following equation
\be \left[\nabla_{\brho}^2 +\varepsilon(\rho)k^2\right]G_k(\brho,\brho')=\delta(\brho-\brho')
\label{GFcyl} \ee
with
\be \varepsilon(\rho)=\left\{
\begin{array}{lll}
n_r^2 & {\rm for}&\rho \leqslant R\,,\\
1& {\rm for} &\rho >R\,.
\end{array}
\right. \ee
Using the angular basis \Eq{chi-n} the GF can be written as
\be G_k(\brho,\brho')=\frac{1}{\sqrt{\rho\rho'}}\sum_m \tilde{G}_m(\rho,\rho';k)
\chi_m(\varphi)\chi_m(\varphi')\,, \ee similar to \Eq{GF-TM}. Note that we redefined here the radial part as
$\tilde{G}_m(\rho,\rho';k)=\sqrt{\rho\rho'}{G}_m(\rho,\rho';k)$ which satisfies
\be
\left[\frac{d^2}{d\rho^2}-\frac{m^2-1/4}{\rho^2}+k^2\varepsilon(\rho)\right]\tilde{G}_m(\rho,\rho';k)=\delta(\rho-\rho')\,.
\ee
Using two linearly independent solutions $f_m(\rho)$ and $g_m(\rho)$ of the corresponding
homogeneous equation which satisfy the asymptotic boundary conditions
$$
\begin{array}{lll}
f_m(\rho)\propto \rho^{m+1/2}&{\rm for}&\rho\to 0\,,\\
g_m(\rho)\propto e^{ik\rho}& {\rm for}&\rho\to \infty\,.
\end{array}
$$
the GF can be expressed as
\be \tilde{G}_m(\rho,\rho';k)=\frac{f_m(\rho_<)g_m(\rho_>)}{W(f_m,g_m)}\,, \ee
in which $\rho_<=\min\{\rho,\rho'\}$, $\rho_>=\max\{\rho,\rho'\}$, and the Wronskian $W(f,g)= f
g'-f'g$\,. For TM polarization, a suitable pair of solutions is given by
\bea f_m(\rho)&=&\sqrt{\rho}\cdot\left\{
\begin{array}{ll}
J_m(n_r\rho)\,, & \rho \leqslant R\,,\\
a_m J_m(\rho) +b_m H_m(\rho)\,,& \rho >R\,,
\end{array}
\right.
\nonumber \\
g_m(\rho)&=&\sqrt{\rho}\cdot\left\{
\begin{array}{ll}
c_m J_m(n_r\rho) +a_m H_m(n_r\rho)\,,& \rho \leqslant R\,,\\
H_m(\rho)\,,& \rho >R\,,
\end{array}
\right. \nonumber \eea
where
\bea a_m(k)&=&\bigl[n_r J_m'(n_r x)H_m(x)-J_m(n_r x)H_m'(x)\bigr]\pi ix/2\,,
\nonumber \\
b_m(k)&=&\bigl[J_m'(x)J_m(n_rx)-n_r J_m(x)J_m'(n_rx)\bigr]\pi ix/2\,,
\nonumber \\
c_m(k)&=&\bigl[H_m'(x)H_m(n_rx)-n_r H_m(x)H_m'(n_r x)\bigr]\pi ix/2 \nonumber \eea
with $x=kR$. The Wronskian is calculated to be
$$
W(f_m,g_m)= 2ia_m(k) /\pi\,=-x D_m(x)\,,
$$
with $D_m(x)$ defined in \Eq{Dfunc}. Inside the cylinder, the GF then takes the form
\bea
\hspace{-1cm}\tilde{G}_m(\rho,\rho';k)&=&\frac{\pi}{2i} \sqrt{\rho\rho'} \biggl[J_m(n_r
k\rho_<)H_m(n_r k\rho_>)
\nonumber\\
&&\left.+\frac{c_m(k)}{a_m(k)}J_m(n_r k\rho_<)J_m(n_r k\rho_>)\right].
\label{GF-Hankel}
\eea
The GF has simple poles $k_n$ in the complex $k$-plane which are the wave vectors of RSs, given by
$a_m(k_n)=0$, an equation equivalent to Eq.\,(\ref{secular2}). The residues
 ${\rm Res}_n$ of the GF at these poles are calculated using
\be r_m(k_n)=\left.\frac{c_m(k)}{\frac{d}{dk} a_m(k)}\right|_{k=k_n}\!\!\!\!\!\!=\frac{2i k_n}{\pi
(n^2_r-1) [k_n R\, J_m(n_r k_nR)]^2}\,. \label{Res} \ee

In addition to the poles, the GF has a cut in the complex $k$-plane along the negative imaginary half-axis.
The cut is due to the Hankel function $H_m(z)$ which describes the field outside the cylinder and
contributes to Eqs.\,(\ref{R-analyt}), (\ref{secular2}) and \Eq{GF-Hankel}, and is not uniquely defined. Indeed, it can
be expressed as\cite{Gradshtein}
\be H_m(z)=J_m(z)+ iN_m(z)\,,
\label{Hankel} \ee
using a multiple-valued Neumann function
\be N_m(z)=\tilde{N}_m(z)+ \frac{2}{\pi}J_m(z) \ln\frac{z}{2}\,, \label{Neumann} \ee
where $\tilde{N}_m(z)=z^m F_m(z^2)$ is a single-valued polynomial\cite{Gradshtein} while
$\ln z$ is a multiple-valued function defined on an infinite number of Riemann sheets. We have verified that
only one such sheet provides the asymptotics $H_m(z)\propto \exp(i z)/\sqrt{z}$ for $z\to\infty$,
which is required for the RS wave functions outside the cylinder to
satisfy the outgoing wave boundary conditions \Eq{BC}. This `physical' sheet has a cut going from
the branch point at $z=0$ to infinity, and the position of this cut is not arbitrary. To find the
cut position let use the symmetry of the RS wave numbers, $k_{-n}=-k_n^\ast$, discussed in Sec.\,\ref{sec:RSE}.
Let us also, using properties of cylindrical functions,\cite{Gradshtein,Bateman53}  bring the secular equation
(\ref{secular2}) to the form
\be J_{m+1}(n_r z)H_{m-1}(z)=J_{m-1}(n_r z)H_{m+1}(z)\,, \label{secular} \ee
in which $z=k_n R$.
We note that if $z=k_nR$ is a complex solution of \Eq{secular}, then
$-z^\ast$ is also a solution of the same equation. We take two equations, one is the conjugate of
\Eq{secular} and the other is \Eq{secular} itself but taken with the argument $-z^\ast$, and add
them up. Substituting there Eqs.\,(\ref{Hankel}) and (\ref{Neumann}) and using the facts that\cite{Gradshtein}
\bea
\bigl[J_m(z)\bigr]^\ast&=&J_m(z^\ast)=(-1)^m J_m(-z^\ast)\,,\nonumber\\
\bigl[\tilde{N}_m(z)\bigr]^\ast&=&\tilde{N}_m(z^\ast)=(-1)^m \tilde{N}_m(-z^\ast)\,,\nonumber \eea
we arrive at the condition
\be
\ln(-z^\ast)-(\ln z)^\ast=\pi i\,, \ee
which is fulfilled, for any $z$, only if $\ln z$ [and consequently $H_m(z)$] has a cut along the negative imaginary half-axis.

Owing to the cut of the Hankel function $H_m(z)$  the GF also
has a cut along the negative imaginary half-axis in the complex $k$-plane, so that on both sides
of the cut $\tilde{G}_m$ takes different values: $\tilde{G}^+_m$ one the right-hand side and
$\tilde{G}^-_m$ one the left-hand side of the cut. The step
$\Delta\tilde{G}_m=\tilde{G}^+_m-\tilde{G}^-_m$ over the cut can be calculated using the
corresponding difference in the Hankel function:
$$
\Delta H_m(z)=H^+_m(z)-H^-_m(z)=4J_m(z)\,.
$$
The result is
\be \Delta\tilde{G}_m(\rho,\rho';k)=\frac{\pi}{2i}\sqrt{\rho\rho'}J_m(n_r
k\rho_<)J_m(n_r k\rho_>)\Delta Q_m(k) \label{deltaG} \ee
where
\be \Delta
Q_m(k)=\left[4+\frac{c_m^+}{a_m^+}-\frac{c_m^-}{a_m^-}\right]=-\left(\frac{4}{\pi
kR}\right)^2\frac{1}{D_m^+(k)D_m^-(k)} \ee
with $D_m(k)$ given by \Eq{Dfunc}.

\begin{figure}[t]
\includegraphics*[width=0.95\columnwidth]{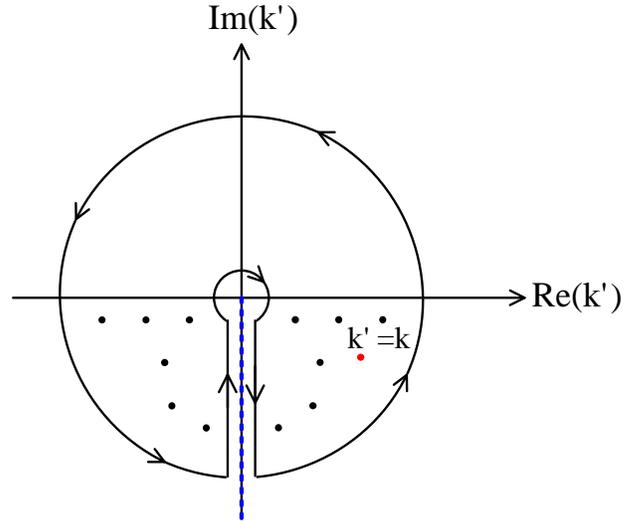}
\caption{Sketch showing the contour of integration in \Eq{integrals} as well as poles (black dots) and the cut (blue dashed line) of the GF in the complex $k'$-plane. An extra pole at $k'=k$ is shown by a red dot. }\label{Fig:contour}
\end{figure}
Let us now use the residue theorem for the function $\tilde{G}_m(\rho,\rho';k')/(k-k')$
integrating it in the complex $k'$-plane along a closed contour consisting of three parts, see Fig.\,\ref{Fig:contour}: A large
counter-clockwise circumference with a radius tending to infinity, two straight lines
circumventing the cut and approaching it from both sides, and a small clockwise circumference
around the origin with a radius tending to zero.  Since the GF behaves as $k^{-2}$ at large values of $k$ and
takes finite values or logarithmically diverges at $k=0$,  both large- and small-circle integrals
vanish, so that the only remaining integrals are those which are taken along the cut:
\bea
&&\oint\frac{\tilde{G}_m(\rho,\rho';k')}{k-k'} d k'= \int^{-i\infty}_0\frac{\tilde{G}^+_m d k'}{k-k'}+\int_{-i\infty}^0\frac{\tilde{G}^-_m d k'}{k-k'}\nonumber\\
&&=2\pi i \sum_n \frac{{\rm Res}_n}{k-k_n}-2\pi i  \tilde{G}_m(\rho,\rho';k)\,. \label{integrals}
\eea
Note that in second part of the above equation we have made use of the residue theorem, expressing
the closed-loop integral in the left-hand side in terms of a sum over residues at all poles inside
the contour. Using \Eq{integrals} the GF can be expressed as
\be
\tilde{G}_m(\rho,\rho';k)=\sum_n \frac{{\rm Res}_n}{k-k_n}+\frac{1}{2\pi i}
\int_{-i\infty}^0\frac{\Delta\tilde{G}_m(\rho,\rho';k') d k'}{k-k'}\,, \label{GF-exp} \ee
which is
a generalization of the Mittag-Leffler theorem. This is used in Sec.\,\ref{sec:application2D} when
applying the RSE to 2D systems with a cut of their GF. The residues ${\rm Res}_n$ of the GF
contributing to \Eq{GF-exp} are calculated as
\be {\rm Res}_n=
\frac{\pi}{2i}\sqrt{\rho\rho'}J_m(n_r k\rho_<)J_m(n_r k\rho_>)r_m(k_n) \,, \label{Res2} \ee
with $r_m(k_n)$ found in \Eq{Res}. Given that the spatial dependence of the GF, as described by
Eqs.\,(\ref{GF-exp}), (\ref{Res2}), and (\ref{deltaG}), is represented by products of the RS wave
functions $R_m(\rho,k_n)$ and their analytic continuations $R_m(\rho,k)$ with $k$-values taken on
the cut, we arrive at the GF in the form of Eqs.\,,(\ref{Gm}) and (\ref{sigma}) which are then used in the RSE.

\section{RSE with a cut}
\label{App:RSEcut}
This section provides some details on how the RSE can be used in practice when the GF has a cut, and in particular, how
the cut discretization, producing cut poles, modifies the linear matrix eigenvalue problem \Eq{RSE}, the central equation of the RSE.

To make our consideration as general as possible we use the 3D version of the Dyson equation
(\ref{Dyson}) as a starting point and  substitute into it spectral representations \Eq{ML2} of the
unperturbed and perturbed GFs. Equating the residues at the perturbed poles (i.e. integrating
along the infinitesimal circumference around each pole) we obtain\cite{Muljarov10}
\be \Epn(\br) = \lefteqn{\sum_n}\int \, \frac{\En(\br)\int \En(\br')\cdot
\Epn(\br')\Delta\varepsilon(\br') d \br'}{2(k_n/\varkappa_\nu-1)} \,, \label{Dyson2} \ee
where both pole and cut contributions are included. The cut has a continuous contribution.
Therefore, in numerical calculation, we discretize the cut representing it by a finite number of
cut poles chosen in an optimum way as described in \Sec{sec:application2D}. For an
arbitrary function $F(k)$,
\be \lefteqn{\sum_n}\int \, F_n\equiv\sum_n F(k_n)+\int_{-i\infty}^0
F(k) \sigma (k) dk \approx\sum_{\bar{n}} F_{\bar{n}} \phi_{\bar{n}} \label{discretiz} \ee
where the combined index $\bar{n}$ is used to denote both real poles $k_n$ and cut poles
$k_\alpha$ simultaneously. The weighting factors $\phi_{\bar{n}}$ come from the weight function
$\sigma(k)$ introduced in \Eq{sigma} and are defined as follows
\be \phi_{\bar{n}}=\left\{
\begin{array}{ll}
\phi_n=1 & {\rm for\ real\ poles},\\
\phi_\alpha & {\rm for\ cut\ poles},\\
\end{array}
\right.
\ee
where
\be \phi_\alpha=\int_{q_\alpha}^{q_{\alpha+1}} \sigma(k)dk\,
\label{phi-al}
\ee
is an integral of the weight function over an interval $[q_\alpha,q_{\alpha+1}]$ within which a
cut pole $k_\alpha$ is chosen. The method of choosing this interval and the positions of the cut
poles specific to this work are described in \Sec{sec:application2D}, by Eqs.\,(\ref{weight}) and
(\ref{k-alpha}). Note that we have dropped here the azimuthal index $m$ for simplicity, however
the discretization of the cut is generally different for different $m$ leading to $m$-dependent
cut poles and their weighting factors $\phi_{\bar{n}}$.

Now, discretizing \Eq{Dyson2}  in accordance with  \Eq{discretiz}, substituting the expansion
\be \Epn(\br) = \lefteqn{\sum_n}\int \, b_{n\nu} \En(\br) \approx \sum_{\bar{n}} \phi_{\bar{n}}
b_{\bar{n}\nu}{\bf E}_{\bar {n}}(\br) \ee
of the perturbed RS wave function into it, and equating
coefficients at the same basis functions, we obtain
\be
b_{\bar{n}\nu}=\frac{1}{2(k_{\bar{n}}/\varkappa_\nu-1)}\sum_{\bar{n}'} \phi_{\bar{n}'}
b_{\bar{n}'\nu} V_{\bar{n}\bar{n}'}\,. \ee
Introducing new expansion coefficients
$$
c_{\bar{n}\nu} = b_{\bar{n}\nu} \sqrt{\frac{k_{\bar{n}}}{\phi_{\bar{n}}\varkappa_\nu}}\,,
$$
the matrix eigenvalue problem takes the form
\be
\sum_{\bar{n}'}\left(\frac{\delta_{{\bar{n}}{\bar{n}}'}}{k_{\bar{n}}}+\frac{V_{{\bar{n}}{\bar{n}}'}}{2}\sqrt{\frac{\phi_{\bar{n}}
\phi_{\bar{n}'}}{k_{\bar{n}} k_{\bar{n}'}}}\right) c_{{\bar{n}}'\nu}=\frac{1}{\varkappa_\nu} c_{{\bar{n}}\nu}\,,
 \label{RSE2}
 \ee
which is a discretized version of \Eq{RSE}. The matrix elements $V_{\bar{n}\bar{n}'}$ are defined
by \Eq{Vnm} with $\bar{n}$ and $\bar{n}'$ numbering both normal and cut poles. They are calculated below in Appendix\,\ref{App:ME} for various types of perturbation.

\section{Matrix elements for various perturbations in 2D}
\label{App:ME}
In this section we give explicit expressions for the matrix elements $V_{\bar{n}\bar{n}'}$  
of the specific perturbations considered in this paper. As a starting point we use the following general formula for the matrix elements of an arbitrary perturbation $\Delta\varepsilon (\rho,\varphi)$ inside the cylinder of radius $R$:
\bea
\hspace{-1cm} V_{{\bar{n}}{\bar{n}}'}&=&\int_0^{2\pi}\chi_m(\varphi)\chi_{m'}(\varphi)d\varphi\nonumber\\
&&\times\int_0^R \Delta\varepsilon (\rho,\varphi)R_m(\rho,k_{\bar{n}}) R_{m'}(\rho,k_{\bar{n}'}) \rho d\rho\,,
\label{ME-cut}
\eea
in which $R_m$ and $\chi_m$ are the eigenfunctions of the homogeneous cylinder given by Eqs.\,(\ref{chi-n})--(\ref{A-norm}).

\subsection{Homogeneous cylinder perturbation}
The homogeneous perturbation \Eq{eps-hom} does not mix different $m$-values. The matrix elements
between RS with the same azimuthal number $m$ are given by the radial overlap integrals
\be
V_{\bar{n}\bar{n}'}=\Delta\epsilon \int_0^R R_m(\rho, k_{\bar{n}})R_m(\rho, k_{\bar{n}'})\rho
d\rho\, \ee
yielding for identical basis states ($\bar{n}=\bar{n}'$)
\be
V_{\bar{n}\bar{n}}=\frac{\Delta\epsilon}{n_r^2-1}\left[1-\frac{J_{m-1}(n_r
k_{\bar{n}}R)J_{m+1}(n_r k_{\bar{n}}R)}{[J_{m}(n_r k_{\bar{n}}R)]^2}\right] \label{V-homogeneous1}
\ee
and for different basis states ($\bar{n}\neq\bar{n}'$)
\bea
V_{\bar{n}\bar{n}'}&=&\frac{\Delta\epsilon}{{n_r^2}-1}\,\frac{2}{n_r
R(k_{\bar{n}}^2-k_{\bar{n}'}^2)} \label{V-homogeneous2}
\\
&&\times\left[k_{\bar{n}'}\frac{J_{m-1}(n_r k_{\bar{n}'}R)}{J_{m}(n_r
k_{\bar{n}'}R)}-k_{\bar{n}}\frac{J_{m-1}(n_r k_{\bar{n}}R)}{J_{m}(n_r
k_{\bar{n}}R)}\right]\,.\nonumber \eea

\subsection{Half-cylinder perturbation}
The most efficient way of calculating the matrix elements of the perturbation \Eq{halfmoon} is to
calculate the angular parts of the integrals analytically and the radial parts numerically. The
matrix elements have the form
\be V_{\bar{n}\bar{n}}=\Delta\epsilon
P_{mm'}Q_{k_{\bar{n}}k_{{\bar{n}'}}}^{mm'}, \label{V-halfmoon1} \ee
in which the angular overlap integrals $P_{mm'}$ are vanishing when taken between modes of
different parity, i.e. between sine and cosine modes, see \Eq{chi-n}, and between same parity
modes corresponding to azimuthal numbers $m$ and $m'$ of different parity. The non-vanishing
integrals are given by %
\bea
P_{mm'}&=&\int_{-\pi/2}^{\pi/2} \!\!\!\!\chi_m(\varphi)\chi_{m'}(\varphi) d\varphi-\int_{\pi/2}^{3\pi/2} \!\!\!\!\chi_m(\varphi)\chi_{m'}(\varphi) d\varphi\nonumber\\
&=&s_m s_{m'} (\psi_{m-m'}\pm\psi_{m+m'}) \label{V-halfmoon2} \eea
with $+$ ($-$) corresponding to cosine (sine) modes and $s_m$ and $\psi_m$ defined as
\bea
s_m=\left\{
\begin{array}{cll}
\pi^{-1/2}&  {\rm for} & m\neq0\,,\\
(2\pi)^{-1/2} & {\rm for}& m=0\,,
\end{array} \right. \nonumber \\
\psi_m=
\bigl[1-(-1)^m\bigr]\,\frac{\sin (m\pi/2)}{m}\,.\nonumber\eea
The radial part of the matrix elements of the perturbation is given by the integrals
\bea
Q_{kk'}^{mm'}&=&\int_0^R R_m(\rho,k)R_{m'}(\rho,k')\rho d\rho \label{Qmm}
\\
&=&\frac{2}{R^2(n_r^2-1)}\,\frac{  \int_0^R J_m(n_r k\rho)J_{m'}(n_r k'\rho)\rho d\rho}{J_m(n_r
kR)J_{m'}(n_r k'R)}\nonumber \eea
which we calculate numerically.
\bigskip

\subsection{Thin-film perturbation}
The matrix elements of the perturbation \Eq{line} are given by the integrals
\be
V_{\bar{n}\bar{n}'}=h\Delta\epsilon \chi_m^2(0)\int_0^R
R_m(\rho,k_{{\bar{n}}})R_{m'}(\rho,k_{{\bar{n}'}}) d\rho\,, \label{V-line} \ee
similar to \Eq{Qmm}, which are calculated numerically.

\subsection{Thin-wire perturbation}

The RSE perturbation matrix elements for this system are calculated by summing $I$ same-strength
delta scatterers on a square grid covering a circle. The perturbation \Eq{point} is thus modeled
by
\be \Delta\varepsilon\approx \Delta \epsilon \frac{\pi b^2}{I}\sum_{i=1}^I
\frac{1}{\rho}\delta(\rho-\rho_i)\delta(\varphi-\varphi_i)\,. \ee
The matrix elements then have the form
\bea V_{\bar{n}\bar{n}'}=\Delta \epsilon \frac{\pi b^2}{I}\sum_{i=1}^I
E_{\bar{n}}(\rho_i,\varphi_i)E_{{\bar{n}}'}(\rho_i,\varphi_i) \label{V-delta} \eea
with $E_{\bar{n}}(\rho,\varphi)$ given by \Eq{E-fact}.

\end{document}